\documentclass[12pt]{revtex4-1}

\usepackage{amsmath,amssymb,graphicx}
\usepackage{epsfig}
\usepackage{textcomp}
\usepackage{color}
\usepackage{cases}
\usepackage{epstopdf}
\usepackage{multirow}
\usepackage{hyperref}
\hypersetup{pdfpagemode=UseNone,colorlinks=true,linktoc=page,pdfborder={0 0 0},linkcolor=blue,citecolor=blue}
\usepackage{geometry}
\begin{document}
\title{Rare events in stochastic populations under bursty reproduction}

\author{Shay Be'er and Michael Assaf\footnote{email address: \texttt{assaf@phys.huji.ac.il}; Corresponding author}}

\affiliation{Racah Institute of Physics, Hebrew University
of Jerusalem, Jerusalem 91904, Israel}


\begin{abstract}

Recently, a first step was made by the authors towards a systematic investigation of the effect of \textit{reaction-step-size noise} -- uncertainty in the step size of the reaction -- on the dynamics of stochastic populations. This was done by investigating the effect of bursty influx on the switching dynamics of stochastic populations. Here we extend this formalism to account for bursty reproduction processes, and improve the accuracy of the formalism to include subleading-order corrections. Bursty reproduction appears in various contexts, where notable examples include bursty viral production from infected cells, and reproduction of mammals involving varying number of offspring. The main question we quantitatively address is how bursty reproduction affects the overall fate of the population. We consider two complementary scenarios: population extinction and population survival; in the former a population gets extinct after maintaining a long-lived metastable state, whereas in the latter a population proliferates despite undergoing a deterministic drift towards extinction. In both models reproduction occurs in bursts, sampled from an arbitrary distribution. In the extinction problem, we show that bursty reproduction broadens the quasi-stationary distribution of population sizes in the metastable state, which results in an exponential \textit{decrease} of the mean time to extinction. In the survival problem, bursty reproduction yields an exponential \textit{increase} in survival probability of the population. Close to the bifurcation limit our analytical results simplify considerably and are shown to depend solely on the mean and variance of the burst-size distribution. Our formalism is demonstrated on several realistic distributions which all compare well with numerical Monte-Carlo simulations.

\end{abstract}

\maketitle

\textbf{--------------------------------------------------------------------------------------------------}
\vspace{-5mm}
\tableofcontents
\textbf{--------------------------------------------------------------------------------------------------}

\section{Introduction}
\label{Introduction}


The mechanism of noise-induced escape often determines the ultimate fate of populations whose dynamics are governed by gain-loss processes. Examples include noise-induced extinction of a population residing in a long-lived metastable state, or noise-induced establishment of a population being initially below some critical threshold, despite having a deterministic drift towards extinction. Here, when the typical population size is large, it is a rare large fluctuation that is responsible for this escape. Such noise-induced escape can be found in various disciplines including  physics, chemistry, biology, ecology, econophysics, and even linguistics, see \textit{e.g.}, Refs. \cite{Horsthemke,ecology2,chem_reac3,stocks3,Elowitz,Pagel1}.

Previous studies regarding noise-induced escape have mainly focused on the role of \textit{demographic}, or intrinsic noise, see \textit{e.g.}, Refs.~\cite{Bartlett,Nisbet,Dykman,KS,explosion,EK,Dykman1,PRE2010,OMII,ours}. Intrinsic noise arises from within the system of interest due to the discrete nature of the population and the inherent stochasticity of the underlying reactions. There exists, however, also \textit{non-demographic} noise, arising from interactions between the system of interest and its noisy environment or from its interaction with other fluctuating systems.

One form of non-demographic noise is \textit{extrinsic} noise, which introduces uncertainty over time in the values of the reaction rates, see \textit{e.g.}, Refs.~\cite{Elowitz,PaulssonI,RajI,RajII,MaoI,MaoII,ecology,KamenevAndBaruch,MickeyElijah}. For example, such time-fluctuating reaction rates can appear in the context of cell biology, e.g., due to cell-to-cell variations~\cite{Elowitz,PaulssonI,RajI,MickeyElijah,OURS-EN}, or in the context of ecology, e.g., due to temperature fluctuations~\cite{ecology}. In these and other cases extrinsic noise may have a dramatic effect on the population's dynamics~\cite{MickeyElijah}.

There is, however, another type of non-demographic noise in the form of \textit{reaction-step-size} noise. This noise introduces uncertainty over time in the step size of the reaction. Previously, only specific examples of this type of noise have been considered, see e.g., Refs.~\cite{KamenevAndBaruch,Swain,Paulsson,AssafRobertsSchulten,SwainI}. Recently, in Ref.~\cite{OURS-BI}, we have developed a formalism to systematically deal with step-size noise, where we have considered a bursty influx process $\emptyset\to kA$, with $k$ being a step-size parameter fluctuating with a given statistics. Importantly, we have shown that bursty influx can exponentially decrease the mean switching time between two metastable states~\cite{OURS-BI}.

In this paper we extend this formalism to allow dealing with bursty reproduction (BR) processes, which are autocatalytic. In this realm we consider two complementary scenarios: extinction of an established population residing in a long lived metastable state, and proliferation (or survival) of a population which has a deterministic drift towards extinction. In the former problem we are interested in quantifying the effect of BR on the mean time to extinction (MTE) and the quasi-stationary distribution (QSD), while in the latter, we seek to unravel the effect of BR on the survival probability (SP).

Our main motivation to study BR comes from the field of virology. Viral production from infected cells can occur either continuously or via a burst~\cite{Krapivsky}. For example, bacterial viruses (bacteriophages) production in \textit{Escherichia coli} has long been recognized to occur in bursts~\cite{1945}. This bursty dynamics is also observed in the production mechanism of the visna virus \cite{Visna}. On the other hand, it has not yet been conclusively determined whether production of the HIV virus occurs continuously or in bursts~\cite{Krapivsky}. Recently, the problem of extinction of the HIV virus with~\cite{Sinitstyn} and without~\cite{Krapivsky} maintaining a long-lived metastable state has been studied, in both the continuous and bursty reproduction variants of the model. The fact that they found a significant difference between the models, motivated us to investigate the generic problem of BR.

Another example which demonstrates the importance of BR comes from the field of conservational ecology. The mean value and variance in number of offspring per birth event (litter size) is  unknown, and can climb to considerably high numbers for small sized organisms \cite{variation_in_litter_size_I,variation_in_litter_size_II,variation_in_litter_size_III,variation_in_litter_size_IV}. Variations in the number of offspring have been shown to decrease the risk of extinction \cite{Ex_litter_size_I,Ex_litter_size_II,Ex_litter_size_III} for established populations, and to increase the survival probability in cases of small populations trying to overcome a strong Allee threshold~\cite{offspring1,offspring2}.

Yet, to the best of our knowledge, the generic problem of bursty reproduction has not been studied analytically in the context of rare events such as extinction or survival. In fact, most studies concerning rare events in stochastic population dynamics often employ an effective description of BR by considering \textit{single-step} birth reactions of the form $A\to2A$ or $2A\to3A$ with a mean reproduction rate of a single offspring per unit time, see \textit{e.g.}, Refs.~\cite{DS,KS,explosion}. In this paper we generalize these studies by considering reactions of the form $mA\to mA+kA$ where $m$ and $k$ are integers, and $k$ is drawn from an arbitrary distribution. We then demonstrate our results on  distributions such as geometric and negative-binomial which capture the main features of realistic offspring number distributions \cite{Ex_litter_size_II}. To quantify the effect of BR we compare our model to its single-step analog with the corresponding rate, such that the mean-field deterministic dynamics is unchanged, and show that BR has a dramatic effect on the stochastic dynamics including rare events. Our analytical calculations are carried out by using both the real- and momentum-space approaches, see below, whose combination allows us to accurately compute the MTE (in the extinction problem), and the SP (when the system possesses a deterministic drift towards extinction).

Here is a plan of the remainder of the paper. In Sec. \ref{Population extinction} we present the population extinction model. Using the real-space approach we derive the QSD of population sizes in the long-lived metastable state, and the MTE (in the leading order). This is followed by several examples for different burst-size distributions (BSDs), and the result close to bifurcation for the MTE, when the metastable population is relatively small. In Sec. \ref{Population survival} we present the population survival model, and calculate the SP. We then provide some examples for different BSDs, and show how the SP considerably simplifies when the initial population is close to the critical population size. The main results are summarized and discussed in Sec. \ref{Summary and discussion}. The Appendix contains a derivation of the sub-leading order correction of the MTE for the extinction model using the momentum-space approach.

\section{Population extinction}
\label{Population extinction}


We consider a stochastic process which is a variant of the Verhulst logistic model with bursty reproduction (BR) rather than regular single-step birth. The microscopic dynamics is defined by the reactions and their corresponding rates
\begin{align}
	\label{Ex-a}
	& A\xrightarrow{\lambda_n}A+kA, \ \ k=0,1,2,\dots,  \ \ \ \ \ \lambda_n=Bn\frac{D(k)}{\langle k\rangle} \nonumber \\ & A\xrightarrow{\mu_n}\emptyset,  \quad\quad\quad\quad\quad\quad\quad\quad\quad\quad\quad  \ \  \mu_n=n+\frac{Bn^2}{N}.
\end{align}
Here $n$ is the size of the population, $B\gtrsim1$ is the average reproduction rate, $N\gg1$ is the typical population size in the long-lived metastable state prior to extinction, $k$ is the offspring number per birth event, and $D(k)$ is an arbitrary, normalized burst size distribution (BSD) with the mean value, $\langle k\rangle$, and variance, $\sigma_k^2$. Furthermore, time is rescaled by the linear death rate. Note that the effective reproduction rate, $B/\langle k\rangle$, was chosen to recover the mean field dynamics of the non-bursty model, in which a single branching reaction $A\to 2A$ that occurs with rate $Bn$ replaces the above BR set of reactions.

At the deterministic level, ignoring demographic noise, the reactions defined in Eq.~(\ref{Ex-a}) yield the following rate equation for the mean population size $\bar{n}(t)$
\begin{equation}
	\label{Ex-b}
	\dot{\bar{n}}=\bar{n}\left(B-1-B\bar{n}/N\right).
\end{equation}
Eq. (\ref{Ex-b}) admits two fixed points: a repelling point at $n=0$, corresponding to extinction and an attracting point at $n_s=N(1-1/B)$, corresponding to the average population size in the long-lived metastable state. At the deterministic level, starting from any initial population size $n_0>0$ the population converges into $n_s$ after a typical time scale $t_r\sim (B-1)^{-1}$, inversely proportional to the rate of linear drift. However, at the stochastic level, $n_s$ is only metastable owing to the existence of an absorbing state at $n=0$. Thus, the long-lived metastable state slowly decays while the extinction probability slowly grows, due to the existence of a small probability flux into the absorbing state, proportional to the inverse of the MTE.

To calculate the QSD of the metastable state and its mean decay time given by the MTE, we have to account for fluctuations. This is done by considering the master equation for $\mathbb{P}_n(t)$ - the probability to find $n$ individuals at time $t$
\begin{align}
	\label{Ex-c}
	\dot{\mathbb{P}}_n = \frac{B}{\langle k\rangle} & \left[\sum_{k=0}^{n-1}D(k)(n-k)\mathbb{P}_{n-k}-\sum_{k=0}^{\infty}D(k)n\mathbb{P}_n\right] \\ + & (n+1)\mathbb{P}_{n+1}-n\mathbb{P}_n+\frac{B}{N}\left[(n+1)^2\mathbb{P}_{n+1}-n^2\mathbb{P}_n\right]. \nonumber
\end{align}
Note that in the first term on the right hand side of Eq. (\ref{Ex-c}) the summation can be formally extended up to infinity since it is assumed that $\mathbb{P}_{n<0}=0$ while the second term is simply $-Bn\mathbb{P}_n/\langle k\rangle$. Multiplying Eq. (\ref{Ex-c}) by $n$ and summing over all values of $n$, we recover the rate equation (\ref{Ex-b}), justifying a-posteriori the rate chosen for the BR process.

\subsection{Quasi-stationary distribution}
\label{Quasi-stationary distribution}


In order to calculate the MTE and QSD we employ the real-space WKB approach in the spirit of Ref.~\cite{PRE2010}. This will allow us to find the quantities of interest within exponential accuracy. In the Appendix we show how to further calculate the pre-exponential correction to the MTE which can be significant, especially close to bifurcation. In extinction problems, provided that $N\gg 1$, at times $t\gg t_r$ when the system has already converged into the long-lived metastable state, the dynamics of $\mathbb{P}_{n}$ is governed by a single time exponent~\cite{Dykman,KS,explosion,EK,PRE2010}
\begin{equation}
	\label{Ex-d}
	\mathbb{P}_{n>0}(t\gg t_r)\simeq\pi_ne^{-t/\tau}, \ \ \ \mathbb{P}_0(t\gg t_r)\simeq 1-e^{-t/\tau},
\end{equation}
where $\tau$ denotes the MTE, and the QSD, $\pi_n$ ($n=1,2,\dots$), is defined as the shape function of the metastable state.

To this end we employ the leading order WKB ansatz, $\pi_n\equiv\pi(Q)\simeq Ae^{-N\mathcal{S}(Q)}$~\cite{Dykman}, where $\mathcal{S}(Q)$  is called the action and $A$ is a normalization factor. Here we have introduced a rescaled coordinate $Q=n/N$, and we assume $n\gg1$ to allow a continuum treatment. Plugging Eq. (\ref{Ex-d}) into the master equation (\ref{Ex-c}) and  neglecting the exponentially small term on the left hand side, we arrive in the leading $\mathcal{O}(N)$ order, at a stationary Hamilton-Jacobi equation $\mathcal{H}(Q,P)=0$, with the Hamiltonian \cite{PRE2010}
\begin{equation}
	\label{Ex-e}
	\mathcal{H}(Q,P)=Q(1+B Q)(e^{-P}-1)+\sum_{k=0}^\infty B Q\frac{D(k)}{\langle k\rangle}(e^{kP}-1).
\end{equation}
Here, we have introduced $P=d\mathcal{S}(Q)/dQ$, the conjugate momentum to the coordinate $Q$, and extended the sum in Eq. (\ref{Ex-c}) up to infinity, see text below Eq. (\ref{Ex-c}).

At this point we pause and revisit the validity of the WKB approach we have used. To perform a Taylor-expansion which transforms the master equation (\ref{Ex-c}) into a Hamilton-Jacobi equation, we have implicitly used the assumption that the step size of the reaction is much smaller than the typical system size~\cite{PRE2010}. However, this requirement is not necessarily satisfied as $D(k)$ may be nonzero even if $k={\cal O}(N)$. Yet, it turns out that it is sufficient to demand that the standard deviation of the BSD, $\sigma_k$, satisfies $\sigma_k\ll N$, so that events of large step sizes which invalidate the perturbation theory contribute a negligible error. Thus, we will henceforth assume that $\sigma_k\ll N$.

We now recast the Hamiltonian~(\ref{Ex-e}) in the following form
\begin{equation}
	\label{Ex-f}
	\mathcal{H}(Q,P)=(1-e^{-P}) B Q \left[e^{P}\mathcal{F}(P)-1/B-Q\right],
\end{equation}
where we have defined $\mathcal{F}(P)=[\sum_{k=0}^\infty e^{kP}D(k)-1]/[\langle k\rangle (e^{P}-1)]$. At this point one usually takes the route of solving the stationary Hamilton-Jacobi equation for $P$ to obtain the phase-space activation trajectory, $P_a(Q)$, that is, the most probable path to extinction. This is followed by an integration of this trajectory to obtain the action and consequently the QSD. Note that the trivial zero-energy trajectories, $Q=0$ and $P=0$, corresponding to the extinction and mean-field lines, respectively, do not contribute to the QSD~\cite{PRE2010}. In our case, the Hamilton-Jacobi equation using~(\ref{Ex-f}) gives rise to a transcendental equation for $P_a(Q)$. However, we can circumvent this difficulty by solving it instead for $Q_a(P)$~\cite{Mendez}. This yields the activation trajectory as a function of the momentum
\begin{equation}
	\label{Ex-g}
	Q_a(P)=e^{P}\mathcal{F}(P)-1/B.
\end{equation}

By definition, see above, the action is given by $\mathcal{S}(Q)=\int^QP_a(Q')dQ'$, where the arbitrary constant can be fixed by putting $\mathcal{S}(Q_s)=0$. This yield
\begin{equation}
	\label{Ex-h}
	\mathcal{S}(Q) =\int_{Q_s}^Q\!\!\!\!P_a(Q')dQ'=\int_0^{P_a(Q)}\!\!\!\!\!\!P'\frac{dQ_a(P')}{dP'}dP' = P_a(Q)e^{P_a(Q)}\mathcal{F}[P_a(Q)]-\int_0^{P_a(Q)}\!\!\!\!\!\!e^{P'}\mathcal{F}(P')dP'.
\end{equation}
Here, we have employed the fact that $P_a(Q_s)=0$ -- the momentum at the fixed point $Q_s=n_s/N$ is zero.

The normalization factor, $A$, is obtained by the condition
\begin{equation}
	\label{Ex-i}
	1=\int_0^\infty\pi(n/N)dn.
\end{equation}
Plugging Eq. (\ref{Ex-i}) with the WKB ansatz $\pi_n\simeq Ae^{-N\mathcal{S}(Q)}$ and expanding the QSD up to second order in its Gaussian vicinity, $A$ can be found as follows:
\begin{equation}
	\label{Ex-j}
	1=A\int_0^\infty \!\! \exp\left[-\frac{(n-n_s)^2}{2\sigma_\mathrm{obs}^2}\right]dn \ \Rightarrow \ A=\frac{1}{\sqrt{2\pi\sigma_\mathrm{obs}^2}},
\end{equation}
where $\sigma_\mathrm{obs}^2=N[d^2\mathcal{S}(Q)/dQ^2|_{Q=Q_s}]^{-1}$ is the observed variance of the QSD. Using the relation $d^2\mathcal{S}(Q)/dQ^2=dP_a(Q)/dQ=[dQ_a(P)/dP]^{-1}$, the observed variance can be explicitly expressed as
\begin{equation}
	\label{Ex-k}
	\sigma_\mathrm{obs}^2=N[1+\mathcal{F}'(0)], \ \ \ \mathcal{F}'(0)=\frac{1}{2}\left[\sigma_k^2/\langle k\rangle+\langle k\rangle-1\right],
\end{equation}
where the prime denotes differentiation with respect to $P$, and we have used the $L'H\hat{o}pital$'s rule to evaluate $\mathcal{F}(0)$ and $\mathcal{F}'(0)$.

What is the meaning of the result~(\ref{Ex-k})? The BSD's mean value and variance for the ``simple" single-step reproduction (SSR) case are $\langle k\rangle=1$ and $\sigma_k^2=0$, respectively, see Table \ref{tab-a}; this results with $\sigma_\mathrm{obs}^2=N$. However, other than the case of SSR, for any arbitrary BSD, $\mathcal{F}'(0)>0$~\cite{OURS-BI} which leads to $\sigma_\mathrm{obs}^2>N$. In other words, we have analytically shown that BR broadens the QSD. One of the main consequences of this is the exponential decrease of the MTE, see below.

\begin {table}[h!]
\footnotesize
\begin{center}
\begin{tabular}{ | p{1.4cm} || p{5.3cm} | p{1.5cm} | p{2.2cm} | p{3.1cm} |}
	\hline
& \multirow{2}{*}{$D(k)$} & \multirow{2}{*}{$\langle k\rangle$} & \multirow{2}{*}{$\sigma_k^2$} & \multirow{2}{*}{$\mathcal{F}(P)$} \\ [0.25cm]
	\hline
\multirow{2}{*}{SSR} & \multirow{2}{*}{$\delta_{k,1}$} & \multirow{2}{*}{$1$} & \multirow{2}{*}{0} & \multirow{2}{*}{$1$} \\ [0.2cm]
	\hline
\multirow{2}{*}{KSR} & \multirow{2}{*}{$\delta_{k,K}$} & \multirow{2}{*}{$K$} & \multirow{2}{*}{0} & \multirow{2}{*}{$\frac{e^{KP}-1}{K(e^P-1)}$} \\ [0.3cm]
	\hline
\multirow{2}{*}{PS} & \multirow{2}{*}{$\frac{\lambda^ke^{-\lambda}}{k!}$} & \multirow{2}{*}{$\lambda$} & \multirow{2}{*}{$\lambda$} & \multirow{2}{*}{$\frac{e^{\lambda(e^P-1)}-1}{\lambda(e^P-1)}$} \\ [0.3cm]
	\hline
\multirow{2}{*}{GE} & \multirow{2}{*}{$\left(\frac{1}{1+b}\right)\left(\frac{b}{1+b}\right)^k$} & \multirow{2}{*}{$b$} & \multirow{2}{*}{$b(b+1)$} & \multirow{2}{*}{$\frac{1}{1+b-be^P}$} \\ [0.3cm]
	\hline
\multirow{2}{*}{NB} & \multirow{2}{*}{${k+a-1\choose k}\left(\frac{1}{1+b}\right)^a\left(\frac{b}{1+b}\right)^k$} & \multirow{2}{*}{$ba$} & \multirow{2}{*}{$ba(b+1)$} & \multirow{2}{*}{$\frac{(1+b-be^P)^{-a}-1}{ba(e^P-1)}$} \\ [0.3cm]
	\hline
\end{tabular}
\caption{Burst size distribution (BSD), its mean value, variance, and corresponding $\mathcal{F}(P)$ for single-step reproduction (SSR), $K$-step reproduction (KSR), Poisson (PS), geometric (GE), and negative-binomial (NB) distributions.}
\label{tab-a}
\end{center}
\end {table}

Putting it all together, the QSD is given by $\pi_n\simeq Ae^{-N\mathcal{S}(Q)}$, with the action~(\ref{Ex-h}) and the normalization factor, given by Eqs.~(\ref{Ex-j}) and (\ref{Ex-k}). Note, that the QSD can only be found explicitly upon providing the BSD, which allows calculating  $\mathcal{F}(P)$ in a straightforward manner. Finally, since the WKB approach formally requires that $Q\gg N^{-1}$, the QSD is valid for $n\gg1$.

\begin{figure}[h!]
\includegraphics[trim = .25in .05in .1in .05in,clip,width=3.4in]{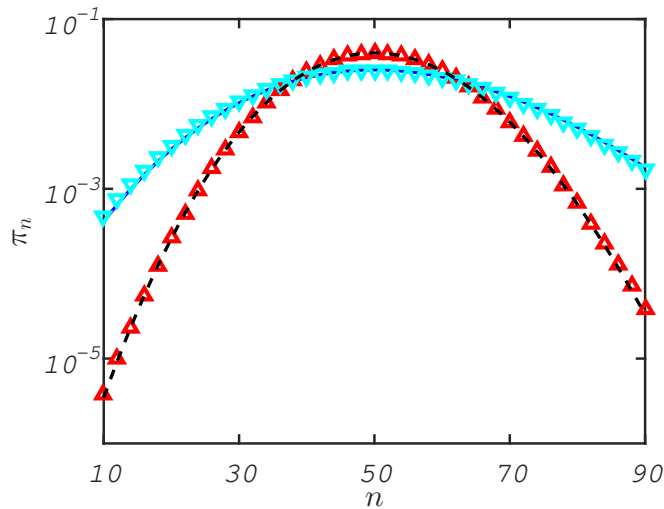}
\caption{(Color online) QSD as a function of the population size. Solid and dashed lines -- theoretical results, see text, for the cases of GE and SSR, respectively, see Table \ref{tab-a}. ($\bigtriangledown$) and ($\bigtriangleup$) markers -- simulation results for the cases of GE and SSR, respectively. The parameters are $b=1.5$, $B=2$, and $N=100$.}
\label{fig-QSD}
\end{figure}

Figure~\ref{fig-QSD} compares between the theoretical and numerical QSD results for the SSR and geometric BSDs as defined in Table~\ref{tab-a}. Our numerical simulations were performed using an extended version of the Gillespie algorithm~\cite{Gillespie} that accounts for BR. An excellent agreement between analytical and numerical results is observed. Furthermore, the broadening of the QSD is well observed, which is a direct consequence of BR, since the mean-field description of the BR and SSR models coincides.

\subsection{Mean time to extinction}
\label{Mean time to extinction}


The MTE is inversely proportional to $\mu_{n=1}\pi_{n=1}$~\cite{PRE2010}. In the leading order, this gives rise to $\tau\simeq e^{N \mathcal{S}(0)}$, where $\mathcal{S}(Q)$ is given by Eq. (\ref{Ex-h}). This result can be simplified. $P_f$, which is called the fluctuational momentum, is defined as $P_a(0)$ -- the value of the momentum at $Q=0$ along the optimal path. It is obtained by solving the transcendent Eq.~(\ref{Ex-g}) with $Q=0$. Substituting the relation found from Eq. (\ref{Ex-g}), $e^{P_f}\mathcal{F}(P_f)=1/B$, into Eq.~(\ref{Ex-h}), and using the definition of $\mathcal{F}(P)$ just below Eq.~(\ref{Ex-f}), we arrive at the MTE:
\begin{align}
	\label{Ex-added}
	& \tau\simeq e^{N \mathcal{S}(0)}, \quad \mathcal{S}(0)=\frac{P_f}{B}-\int_0^{P_f}e^{P'}\mathcal{F}(P')dP' \\
& = \frac{P_f}{B}+\frac{1}{\langle k\rangle}\left\{\ln(1-R)+\gamma+\sum_{m=1}^\infty D(m-1)\left[\frac{R^m}{m}\ _2F_1(1,m,m+1;R)+\psi(m)\right]\right\}. \nonumber
\end{align}
Here $_2F_1(a,b,c;z)=\sum_{l=0}^\infty(a)_l(b)_l/(c)_lz^l/l!$ is the Gaussian hypergeometric function, $\gamma\simeq 0.577$ is Euler's constant, $\psi(z)=\Gamma'(z)/\Gamma(z)$ is the digamma function, and we have defined $R=e^{P_f}$.

Calculating the pre-exponential correction to the MTE is more involved. As a first step, it requires deriving the WKB solution up to sub-leading order. This solution has to be then matched to a recursive solution valid at small population sizes~\cite{PRE2010}. However, here this approach breaks down as, in general, the method of calculating the recursive solution is invalid in the presence of BR. Indeed, this method assumes that for small population sizes the dominant reactions are linear in the population size, and thus, nonlinear terms are neglected. Yet, in the presence of BR, as the step size increases, the rate of the reactions may decrease even below the rates of the nonlinear reactions which were neglected. As a result, unless we assume a very narrow BSD, employing this approach here may lead to an invalid recursive solution, which ultimately prevents the calculation of the subleading-order correction to the MTE.

In the Appendix we take a different route and calculate the MTE including sub-leading order corrections, by employing the alternative momentum-space approach, based on the generating function technique in conjunction with the spectral formalism \cite{spectral_I,spectral_II,spectral_III,MS}. This calculation yields
\begin{equation}	
	\label{Ex-m}
	\tau=\frac{1}{(B-1)(e^{-P_f}-1)} \ \sqrt{\frac{2\pi}{N|Q'_a(P_f)|}}e^{N\mathcal{S}(0)},
\end{equation}
where prime denotes differentiation with respect to $P$, $Q_a(P)$ is given by Eq. (\ref{Ex-g}), $P_f$ is obtained by solving the equation $Q_a(P_f)=0$, and $\mathcal{S}(0)$ is given by Eq. (\ref{Ex-added}). Note that, the MTE is well behaved and is always positive as $B>1$ and $P_f<0$. Eq. (\ref{Ex-m}) is one of the main results of this paper.

\subsection{Examples}
\label{Ex-Examples}


We now illustrate our theory by calculating the MTE (\ref{Ex-m}) for three representative BSDs. In the simple case of the SSR, see Table \ref{tab-a}, Eq. (\ref{Ex-m}) reduces to
\begin{equation}
	\label{Ex-n}
	\tau_{_{\mathrm{SSR}}}=\frac{\sqrt{B}}{(B-1)^2} \ \sqrt{\frac{2\pi}{N}}\exp\left\{\frac{N}{B}\left[B-1-\ln(B)\right]\right\},
\end{equation}
which naturally coincides with the MTE for the Verhulst model, see \textit{e.g.}, Ref.~\cite{PRE2010}.

Next, we generalize the above result and consider the case of a constant step-size $K$ (where $K$ is a positive integer), see Table \ref{tab-a}. The MTE in this case is given by
\begin{align}
	\label{Ex-s}
	& \tau_{_{\mathrm{KSR}}}=\frac{1}{B-1} \ \sqrt{\frac{2\pi R K}{N|1+KR^{K+1}-(K+1)R^K|}} \\ & \times\exp\left\{\frac{NP_f}{B}+\frac{N}{K}\left[\ln(1-R)+\gamma+\frac{R^{K+1}}{K+1}\ _2F_1(1,K+1,K+2;R)+\psi(K+1)\right]\right\}\!\!, \nonumber
\end{align}
where to remind the reader, $R=e^{P_f}$, and $P_f$ is found numerically by solving $Q_a(P_f)=0$.

Finally, in the case of geometric BSD, the MTE is given by
\begin{equation}
	\label{Ex-r}
	\tau_{_{\mathrm{GE}}}=\frac{B(1+b)}{(B-1)^2} \ \sqrt{\frac{2\pi}{N(B+b)}}\exp\left\{\frac{N}{B}\left\{\ln\left(\frac{1+b}{B+b}\right)+\frac{B}{b}\ln\left[\frac{(1+b)B}{B+b}\right]\right\}\!\!\right\}\!\!,
\end{equation}
where $b$ is the mean of the geometric BSD, see Table \ref{tab-a}.

\begin{figure}[h]
\includegraphics[trim = .05in .05in .15in .5in,clip,width=3.4in]{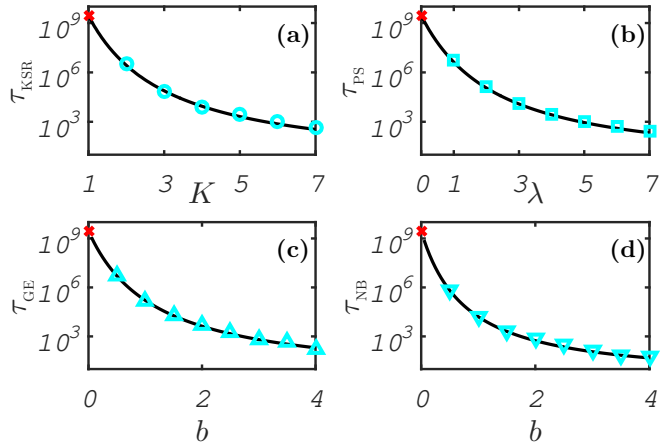}
\caption{(Color online) MTE for various BSDs as a function of their characteristic parameters, see Table \ref{tab-a}. Solid lines -- theoretical results given by Eq.~(\ref{Ex-m}) [where for panels (a) and (c) the results are  explicitly given by Eqs. (\ref{Ex-s}) and (\ref{Ex-r}), respectively]. ($\circ$), ($\Box$), ($\bigtriangledown$), and ($\bigtriangleup$) markers - simulation results for the cases of KSR, PS, GE, and NB, respectively. (x) marker - theoretical value for the case of SSR~(\ref{Ex-n}).  Note, that $\tau_{_{\mathrm{SSR}}}=\lim_{_{K\to1}}\tau_{_{\mathrm{KSR}}}\!=\lim_{_{\lambda\to0}}\tau_{_{\mathrm{PS}}}\!=\lim_{_{b\to0}}\tau_{_{\mathrm{GE}}}\!=\lim_{_{b\to0}}\tau_{_{\mathrm{NB}}}$ which stems from the limit $\delta_{_{k,1}}=\lim_{_{K\to1}}D(k)/\langle k\rangle_{_{\mathrm{KSR}}}\!=\lim_{_{\lambda\to0}}D(k)/\langle k\rangle_{_{\mathrm{PS}}}\!=\lim_{_{b\to0}}D(k)/\langle k\rangle_{_{\mathrm{GE}}}\!=\lim_{_{b\to0}}D(k)/\langle k\rangle_{_{\mathrm{NB}}}$. Here, the parameters are $a=2$, $B=2$, and $N=150$.
}
\label{fig-ResExt}
\end{figure}

Figure~\ref{fig-ResExt} compares between the theoretical and numerical MTEs for various BSDs as a function of their characteristic parameters, see Table \ref{tab-a}, and excellent agreement is observed. For all BSDs, an exponential reduction in the MTE (compared to the non-bursty case) is demonstrated, due to the broadening of the corresponding QSDs via the mechanism of BR. As the deterministic description of BR remains unchanged compared to the case of SSR, this reduction is an exclusive effect of BR.

\subsection{Bifurcation limit}
\label{Bifurcation limit}


Close to bifurcation, the general result for the MTE [Eq.~(\ref{Ex-m})] becomes considerably simpler, and reduces to an expression which solely depends on the mean and variance of the BSD. In the bifurcation limit, the attracting fixed point $Q_s=1-1/B$ is assumed to be close to $Q=0$ such that $Q_s\equiv\delta\ll 1$. As a result, we have $B\simeq 1+\delta$.

We can now use the smallness of $\delta$ to find $P_f$ explicitly. Using Eq.~(\ref{Ex-g}), and putting $P=P_f$ (which causes the left hand side to vanish), we expand the right hand side in powers of $P_f\ll 1$. Using the fact that $B=1+\delta$, we arrive at $P_f\simeq-\delta/[1+\mathcal{F}'(0)]$, thus justifying a-posteriori our assumption $P_f\ll 1$. Having found $P_f$ as a function of $\delta$, we can now expand the action (\ref{Ex-added}) in powers of $\delta$, yielding $\mathcal{S}(0)\simeq \delta^2/[2(1+\mathcal{F}'(0))]$. This allows us to calculate the MTE in the bifurcation limit, up to subleading-order corrections, by using Eq. (\ref{Ex-m}) with the values of $B$ and $P_f$ close to bifurcation. Doing so, we arrive at
\begin{equation}
	\label{Ex-q}
	\tau^{\mathrm{B}}=\frac{\sqrt{2\pi[1+\mathcal{F}'(0)]}}{\sqrt{N}\delta^2}\exp\left\{\frac{N\delta^2}{2[1+\mathcal{F}'(0)]}\right\}.
\end{equation}
To remind the reader, $\mathcal{F}'(0)=\frac{1}{2}\left[\sigma_k^2/\langle k\rangle+\langle k\rangle-1\right]>0$, and thus the MTE is exponentially decreased compared to the SSR case, for which $\mathcal{F}'(0)=0$. Note that in the case of SSR, Eq. (\ref{Ex-q}) coincides with the result for the simple Verhulst model close to bifurcation, as appears in Ref.~\cite{PRE2010}.

What is the validity region of the result~(\ref{Ex-q})? On the one hand, the WKB approach requires that $N\delta^2\gg 1$. On the other hand, in the course of the derivation above, we have neglected terms on the order of $N\delta^3$ in the action. Thus, to avoid excess of accuracy by keeping the pre-exponential factor,  Eq.~(\ref{Ex-q}) is valid as long as $N^{-1/2}\ll\delta\ll N^{-1/3}$.

\begin{figure}[h!]
\includegraphics[trim = .1in 1.1in .4in .3in,clip,width=3.4in]{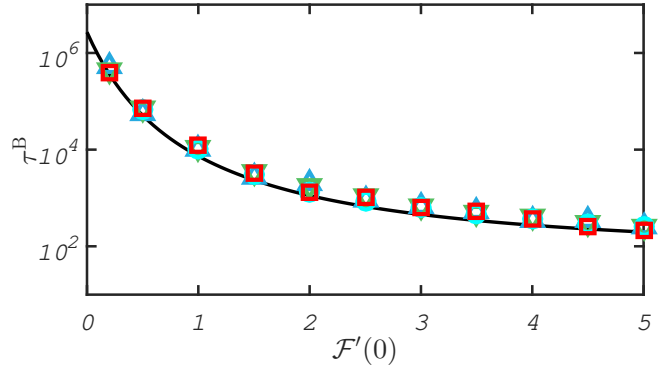}
\caption{(Color online) MTE close to bifurcation as a function of $\mathcal{F}'(0)$. Solid line - theoretical result, given by Eq.~(\ref{Ex-q}). ($\circ$), ($\Box$), ($\bigtriangledown$), and ($\bigtriangleup$) markers - simulation results for the cases of KSR, PS, GE, and NB, respectively, see Table \ref{tab-a}. The parameters are $a=2$, $N=10^4$, and $B=1.053$ such that $\delta=0.05$.}
\label{fig-BfrExt}
\end{figure}

Figure \ref{fig-BfrExt} shows a collapse of simulation results for various BSDs onto the theoretical  result in the bifurcation limit, given by Eq.~(\ref{Ex-q}). This indicates a universal scaling, for any BSD, of the exponential reduction in the MTE as function of $\mathcal{F}'(0)$.

The bifurcation limit allows for an additional simplification of the problem. It turns out that the MTE corresponding to the general model, which involves an infinite set of birth reactions $A\to (k+1)A$ with rates proportional to $D(k)$, can be effectively described by a model involving only a single $K$-step birth reaction $A\to A+KA$. Close to bifurcation, the burstiness is entirely captured by $\mathcal{F}'(0)$, see Eq. (\ref{Ex-q}). Thus, the effective step size of the analogues KSR model is obtained by demanding that $\mathcal{F}'(0)|_{_{KSR}}=(1/2)(K-1)$ coincide with the corresponding $\mathcal{F}'(0)$ of the general model. This leads to
\begin{equation}
	\label{Ex-t}
	K=\langle k\rangle+\sigma_k^2/\langle k\rangle.
\end{equation}

\begin{figure}[h!]
\includegraphics[trim = .2in 1.5in .1in .1in,clip,width=4in]{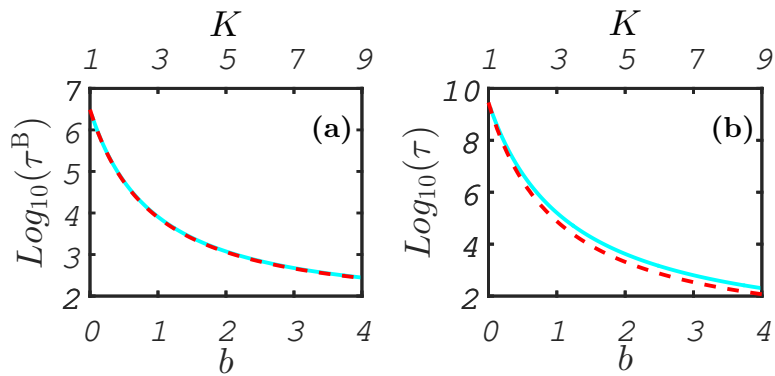}
\caption{(Color online) Panel (a): MTE close to bifurcation given by Eq.~(\ref{Ex-q}). The parameters are $a=2$, $\delta=0.05$, and $N=10^4$. Panel (b): MTE far from bifurcation, given by Eq.~(\ref{Ex-m}). The parameters are $a=2$, $\delta=0.5$, and $N=150$. The solid lines represent the MTE as a function of the characteristic parameter $b$ of the geometric BSD, while the dashed lines represent the MTE as a function of the effective $K$ given by $K=1+2b$ according to Eq.~(\ref{Ex-t}).}
\label{fig-GEByK}
\end{figure}

In Fig.~\ref{fig-GEByK}{\color{blue}(a)}, we demonstrate this coincidence between the MTEs of the geometric BSD and the effective KSR models, close to bifurcation, by adjusting the value of $K$ according to Eq.~(\ref{Ex-t}). Furthermore, our numerical simulations indicate that this result extends beyond the bifurcation limit, as demonstrated by Fig.~\ref{fig-GEByK}{\color{blue}(b)}.

\section{Population survival}
\label{Population survival}


Here we consider a runaway model of a stochastic population including bursty pair reproduction. The microscopic dynamics is defined by the reactions and their corresponding rates
\begin{align}
	\label{Su-a}
	2 & A\xrightarrow{\lambda_n}2A+kA, \ \ k=0,1,2,\dots,  \ \ \ \ \ \lambda_n=\frac{Bn(n-1)}{N}\frac{D(k)}{\langle k\rangle} \nonumber \\ & A\xrightarrow{\mu_n}\emptyset,  \quad\quad\quad\quad\quad\quad\quad\quad\quad\quad\quad  \ \ \ \  \mu_n=n,
\end{align}
where all quantities are defined as in Sec. \ref{Population extinction}, time is rescaled by the linear death rate, and the effective reproduction rate, $B/(N\langle k\rangle)$, was chosen to recover the mean-field dynamics of the non-bursty model. In the deterministic picture, this model is described by the rate equation for the mean population size $\bar{n}(t)$
\begin{equation}
	\label{Su-b}
	\dot{\bar{n}}=\bar{n}\left[\frac{B}{N}(\bar{n}-1)-1\right].
\end{equation}
Eq.~(\ref{Su-b}) admits two fixed points: an attracting fixed point at $n=0$ corresponding to extinction, and a repelling fixed point $n_u=N/B+1\simeq N/B$ corresponding to the critical population size, above which the population enters a state of an unlimited growth (sometimes refereed to as runaway or explosion). The emergence of the critical population size is a result of the strong Allee effect which describes a decrease in the growth rate per capita as the population grows in size~\cite{Allee}.

Note, that this model is a simplified version of the more realistic model with an additional attracting fixed point $n_3>n_u$ corresponding to population establishment. Yet, it can be shown that when $n_3$ is sufficiently distant from $n_u$, the survival and establishment probabilities of the two models coincide in the leading order~\cite{ours}.

In contrast of the extinction problem, see Sec. \ref{Population extinction}, this problem does \textit{not} feature metastability. Instead, for a given initial population $n_0$ the population will deterministically flow towards extinction for $n_0<n_u$, whereas for $n_0>n_u$ it will deterministically proliferate. However, even if the initial population satisfies $n_0<n_u$, it still has a nonzero probability to avoid extinction and to survive, by undergoing a large fluctuation and overcoming the potential barrier at $n_u$. We are interested to calculate $\mathcal{P}(n_0)$ -- the survival probability (SP), starting from $n_0$ individuals.

Our starting point is the master equation for $\mathbb{P}_n(t)$ - the probability to find $n$ individuals at time $t$
\begin{equation}
	\label{Su-c}
	\dot{\mathbb{P}}_n=\frac{B}{N\langle k\rangle}\left[\sum_{k=0}^{n-2}D(k)(n-k)(n-k-1)\mathbb{P}_{n-k}-\sum_{k=0}^{\infty}D(k)n(n-1)\mathbb{P}_n\right]+(n+1)\mathbb{P}_{n+1}-n\mathbb{P}_n.\end{equation}
Note that in the first term on the right hand side of Eq. (\ref{Su-c}) the summation can be formally extended up to infinity since it is assumed that $\mathbb{P}_{n<0}=0$, while the second term is simply $-Bn(n-1)\mathbb{P}_n/(N \langle k\rangle)$.

\subsection{Survival probability}
\label{Survival probability}


The survival probability $\mathcal{P}(n_0)$ can be explicitly calculated by solving the recursion equation for $\mathcal{P}(n_0)$ which is related to the backward master equation~\cite{Gardiner,Krapivsky,EPL,MobiliaII}. However, this equation is only solvable in particular cases and, in general, the solution is highly cumbersome.

Instead, here we follow a different route and consider a modified problem obtained by supplementing the original problem with a reflecting boundary at $n=n_0$. Starting from an initial population $n_0<n_u$, this gives rise to a long-lived metastable state peaked at $n=n_0$ due to the fact that there exists a deterministic drift from $n_u$ towards $n_0$. This metastable state, however, slowly decays due to a slow leakage of probability through the repelling fixed point $n_u$, which eventually results in population escape. Importantly, provided that $n_0$ is not too close to $n_u$, see below, it can be shown that the escape rate in the modified problem equals in the leading order to the SP in the original problem starting from $n_0$ individuals~\cite{Mobilia}.
As a result, we now consider henceforth the modified problem with a reflecting wall at $n=n_0$.

\begin{figure}[h!]
\includegraphics[trim = .2in .1in .1in 0in,clip,width=3.2in]{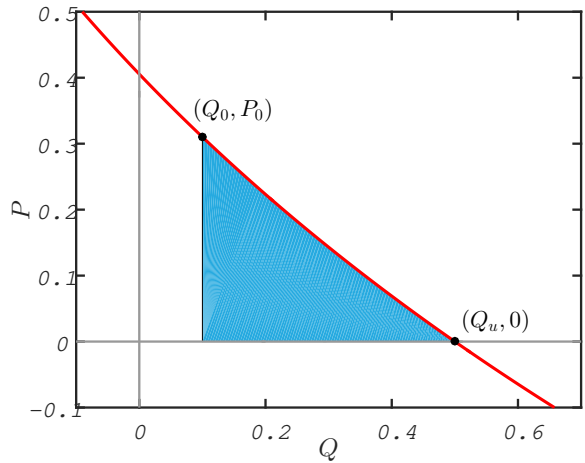}
\caption{(Color online) Shown is the phase-space of the escape problem. Solid line - a characteristic activation trajectory, $Q_a(P)$ given by Eq.~(\ref{Su-e}), for the case of geometric BSD, see Table~\ref{tab-a}. Shaded area - the accumulated action given by Eq.~(\ref{Su-g}). The parameters are $B=b=2$ and $Q_0=0.1$.}
\label{fig-PhsSpc}
\end{figure}

As the modified problem possesses a long-lived metastable state at $n=n_0$, we can employ the leading order WKB ansatz, $\pi(Q)\simeq e^{-N\mathcal{S}(Q)}$. Plugging this ansatz into Eq. (\ref{Su-c}) and repeating the calculations preformed in Sec. \ref{Quasi-stationary distribution}, we arrive at a Hamilton-Jacobi equation, $\mathcal{H}(Q,P)=0$, with the Hamiltonian
\begin{equation}
	\label{Su-d}
	\mathcal{H}(Q,P)=\left(1-e^{-P}\right)BQ\left[Qe^{P}\mathcal{F}(P)-1/B\right],
\end{equation}
where $\mathcal{F}(P)$ is defined in Sec. \ref{Quasi-stationary distribution}. From Hamiltonian (\ref{Su-d}), we can write a transcendental equation for the activation trajectory $P_a(Q)$. However, similarly as before, we rather calculate $Q_a(P)$ which can be done explicitly, yielding
\begin{equation}
	\label{Su-e}
	Q_a(P)=\frac{1}{Be^{P}\mathcal{F}(P)}.
\end{equation}
Figure \ref{fig-PhsSpc} shows the phase-space of the problem and the activation trajectory.

Having found $Q_a(P)$, the action can be found as follows
\begin{equation}
	\label{Su-f}
	\mathcal{S}(Q)=\int_{Q_0}^QP_a(Q')dQ'=P_a(Q)Q -P_0Q_0-\int_{P_0}^{P_a(Q)}Q_a(P')dP'.
\end{equation}
Here, $Q_0=n_0/N$ and $P_0$ is defined as $P_0=P_a(Q_0)$. As a result, the SP of a population of initial size $n_0<n_u$ which is given by the escape rate in the modified problem having a reflecting wall at $n=n_0$, is given by
\begin{equation}
	\label{Su-g}
	\mathcal{P}(n_0)=e^{-N\Delta\mathcal{S}},\quad \Delta\mathcal{S}=\mathcal{S}(n_u/N)-\mathcal{S}(n_0/N)=-P_0n_0/N+\int_0^{P_0}Q_a(P')dP'.
\end{equation}
where we have used the equality $P_a(n_u/N)=0$. Finally, the WKB approach requires that the accumulated action be large. That is $n_0$ must be sufficiently distant form $n_u$ such that  $N\Delta\mathcal{S}\gg1$~\cite{ours}.

\begin{figure}[h!]
\includegraphics[trim = .2in .1in .1in 0in,clip,width=3.4in]{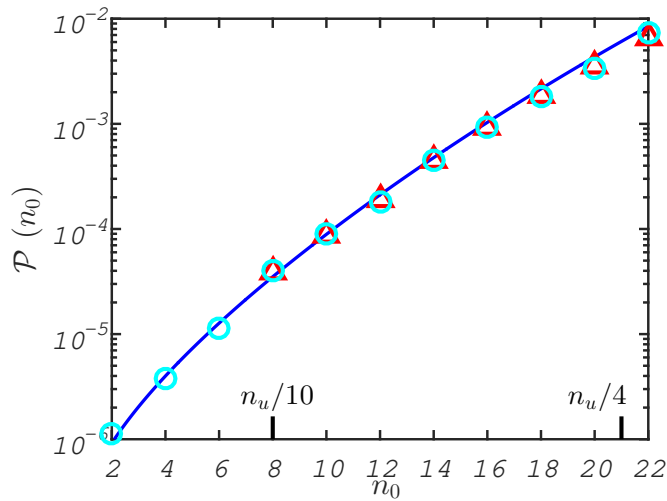}
\caption{(Color online) SP as a function of the initial population size for the case of geometric BSD, see Table \ref{tab-a}. Solid line - theoretical result, Eq. (\ref{Su-m}), multiplied by $Q_0$, see text. ($\bigtriangleup$) marker - simulation results corresponding to the original problem; here, the SP is obtained by dividing the number of iterations which result with an escape by the total number of iterations. ($\circ$) marker - simulation results corresponding to the modified problem with a reflecting wall at $n_0$ possessing metastability. Here we compute the mean escape rate which is proportional to the SP, see text. The theoretical and numerical results for the mean escape rate are normalized by constant factors to fit the SP simulation results. The parameters are $b=2$, $B=1.2$, and $N=100$.}
\label{fig-WallVSNonWall}
\end{figure}

Note, that Eq.~(\ref{Su-g}), which includes only the leading-order contribution, ignores the boundary condition at $n_0=0$, $\mathcal{P}(n_0=0)=0$. Thus, in order to satisfy this boundary condition, we multiply the result~(\ref{Su-g}) by $Q_0$, which excellently agrees with numerical simulations, see Fig.~\ref{fig-WallVSNonWall}.

\subsection{Examples}
\label{Su-Examples}


We now illustrate our theory by calculating the SP (\ref{Su-g}) for the same representative BSDs that were considered in Sec. \ref{Ex-Examples}. From Eq. (\ref{Su-g}) we have $\ln[\mathcal{P}(n_0)]=-N\Delta\mathcal{S}$ with $\Delta\mathcal{S}$ given by
\begin{align}
	\label{Su-m}
	& \Delta\mathcal{S}_{_{\mathrm{SSR}}}=\frac{1}{B}-\frac{n_0}{N}\left[1-\ln\left(Bn_0/N\right)\right], \nonumber \\
	& \Delta\mathcal{S}_{_{\mathrm{KSR}}}=-P_0\frac{n_0}{N} +\frac{1}{B}\left[\ln(1-e^{KP_0})+\gamma+\psi(-1/K)\right] \nonumber \\ & \quad\quad\quad\quad  -\frac{K}{B}\left[P_0+e^{-P_0} \!\!\ _2F_1(1,-1/K,1-1/K;e^{KP_0})\right], \nonumber \\
	& \Delta\mathcal{S}_{_{\mathrm{GE}}}=\frac{1}{B}-\frac{n_0}{N}-\left(\frac{b}{B}+\frac{n_0}{N}\right)\ln\left(\frac{1+b}{b+Bn_0/N}\right),
\end{align}
where all quantities are defined as in Sec. \ref{Ex-Examples}. For the case of K-step reproduction, $P_0$ is determined by a numerical solution of Eq. (\ref{Su-e}).

The equivalence between the original problem without metastability and the modified problem with metastability is numerically demonstrated in Fig.~\ref{fig-WallVSNonWall}. The figure shows three curves for the case of geometric BSD: the numerical SP in the original problem as a function of the initial population size, and the (normalized) theoretical and numerical results for the mean escape rate in the modified problem as a function of the position of the  reflecting wall. Indeed, an excellent agreement between the SP in the original problem and the (normalized) escape rate in the modified problem is observed. As stated above, this equivalence between the problems can be proved theoretically, see \textit{e.g.}, Refs.~\cite{Mobilia}.

\begin{figure}[h!]
\includegraphics[trim = .1in .25in 0in .3in,clip,width=3.4in]{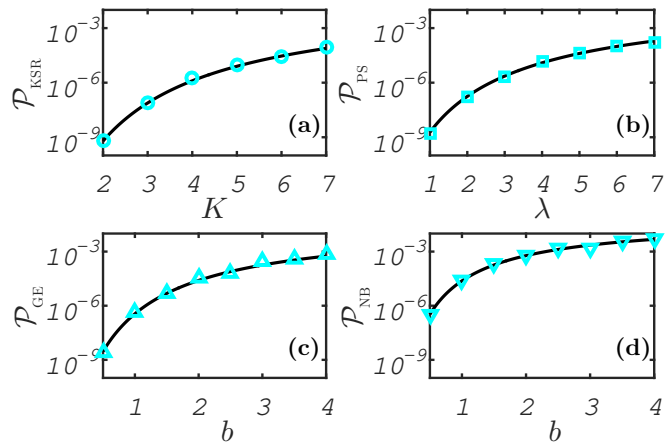}
\caption{(Color online) SP for various BSDs, calculated using the modified problem, see text, as a function of their characteristic parameters, see Table \ref{tab-a}. Solid lines - theoretical results given by Eq.~(\ref{Su-g}) [where for panels (a) and (c) the results are specifically given by Eq.~(\ref{Su-m})]. ($\circ$), ($\Box$), ($\bigtriangledown$), and ($\bigtriangleup$) markers - simulation results corresponding for the cases of KSR, PS, GE, and NB, respectively. The parameters are $a=3$, $B=2$, $n_0=20$, and $N=150$.}
\label{fig-ResSur}
\end{figure}

Having justified our usage of the modified problem, Fig.~\ref{fig-ResSur} shows an excellent agreement between the theoretical and numerical escape rates in the modified problem for various BSDs, see Table \ref{tab-a}. To increase efficiency, we have simulated the modified instead of the original problem, but we have normalized the escape rates to equal the SP by comparing to a single SP simulation of the original problem for each BSD. For all BSDs, an exponential increase in SP is observed. Similarly as in the case of Sec.~\ref{Ex-Examples}, this increase is an exclusive effect of BR as the deterministic description remains unchanged compared to SSR.

\subsection{Near-threshold initial-population limit}
\label{Near-threshold initial-population limit}


We now consider the case in which $n_0$ is close to $n_u$ (but not too close, see below), namely where the initial population is close to the critical population size, above which the population enters a state of an unlimited growth. In this case, the SP considerably simplifies compared to the general result (\ref{Su-g}). Similarly to the bifurcation limit of the extinction problem, see Sec. \ref{Bifurcation limit}, the result here is reduced to an expression which solely depends on the mean and variance of the BSD, see below.

We define the near-threshold initial-population limit as $Q_u-Q_0\equiv\delta\ll 1$ where $\delta$ is the distance to the threshold. This determines the relation
\begin{equation}
	\label{Su-h}
	Q_0=\frac{1}{B}-\delta.
\end{equation}
Note that this limit is \textit{not} a bifurcation limit as the point $Q_0$ is not a fixed point. In fact, the distance between $Q=0$ and $Q_u$ is independent on $\delta$.

Assuming a-priori that the momentum is small we define $P=\tilde{P}\delta$ where $\tilde{P}=\mathcal{O}(1)$. Expanding $\mathcal{F}(P)$ in $\delta\ll 1$  up to leading order we find $\mathcal{F}(P)\simeq1+\mathcal{F}'(0)\tilde{P}\delta$ where prime denotes differentiation with respect to $P$. Substituting this into Eq.~(\ref{Su-e}) and expanding in powers of $\delta$ up to leading order we find
\begin{equation}
	\label{Su-i}
	Q_a(\tilde{P})=\frac{1}{B}\left\{1-[1+\mathcal{F}'(0)]\tilde{P}\delta\right\}.
\end{equation}
Demanding that Eq.~(\ref{Su-i}) evaluated at $P=P_0$ be equal to Eq.~(\ref{Su-h}) we arrive at
\begin{equation}
	\label{Su-j}
	\tilde{P}_0\simeq\frac{B}{1+\mathcal{F}'(0)},
\end{equation}
which justifies our a-priori assumption that $P={\cal O}(\delta)$.

We are now in a position to determine the SP when $n_0$ is close to $n_u$. Changing the integration variable in Eq.~(\ref{Su-g}) and using Eqs. (\ref{Su-h}), (\ref{Su-i}), and (\ref{Su-j})  we finally arrive at
\begin{equation}
	\label{Su-k}
	\mathcal{P}^{\mathrm{B}}(n_0)=e^{-N\Delta\mathcal{S}^{\mathrm{B}}},\quad \Delta\mathcal{S}^{\mathrm{B}}=\frac{B\delta^2}{2[1+\mathcal{F}'(0)]}.
\end{equation}
To remind the reader $\mathcal{F}'(0)=\frac{1}{2}\left[\sigma_k^2/\langle k\rangle+\langle k\rangle-1\right]>0$ and equals to zero only for the SSR case. Note that this result is valid as long as $N\Delta\mathcal{S}\gg1$, thus $\delta$ cannot be too small and must satisfy $N^{-1/2}\ll\delta\ll 1$.

\section{Summary and discussion}
\label{Summary and discussion}


In this paper we have studied the effect of bursty reproduction on the dynamics of stochastic populations including rare events. We have considered two complementary scenarios: population extinction from a long-lived metastable state, and population survival against a deterministic force. In the former we have calculated the quasi-stationary distribution of the population sizes prior to extinction and the mean time to extinction, while in the latter scenario we have calculated the survival probability of a population despite having a deterministic drift towards extinction.  In both scenarios we have presented analytical results for generic burst-size distribution (BSD) and found explicit results for several representative examples. Importantly, we have demonstrated that bursty reproduction can exponentially decrease the mean time to extinction and exponentially increase the survival probability. This occurs due to the broadening of the corresponding probability distribution of population sizes prior to escape. In particular, we have shown that the results considerably simplify when the metastable state/initial population size are close to the corresponding unstable fixed points, in the extinction and survival problems, respectively. In these regimes, we have shown that the results solely depend on the mean and variance of the BSD.

In a previous study, we have considered reaction-step-size noise in the form of bursty influx of individuals, and calculated the mean escape time within exponential accuracy. In this work we have extended the formalism to allow treating bursty autocatalytic processes which depend on the population size. Furthermore, we have shown how the subleading-order pre-exponential corrections to MTE and QSD can be calculated by employing both the real-space approach, and also the momentum-space approach in conjunction with the spectral formalism.

Finally, the analytical results we have derived here for the dynamics of birth-death processes including rare events, under generic bursty reproduction, may be of high importance in a variety of systems, including population biology and viral dynamics, where bursty reprodction constitutes a key mechanism in the long-time behavior of such systems.

\section*{Acknowledgments}

\noindent
We would like to thank Yonatan Friedman for a useful discussion.

\section*{Appendix - Sub-leading order calculations}
\label{Appendix A}
\renewcommand{\theequation}{A\arabic{equation}}
\setcounter{equation}{0}


Here we derive the sub-leading order correction to the MTE. We use the momentum-space approach by employing the generating function technique~\cite{Gardiner,Elgart} in conjunction with the spectral formalism \cite{spectral_I,spectral_II,spectral_III,MS}.

The probability generating function of the population is defined as \cite{Gardiner}
\begin{equation}
	\label{Ap-b}
	G=\sum_{n=0}^\infty p^n \mathbb{P}_n,
\end{equation}
where $p$ is an auxiliary variable that will play the role of the momentum, see below. Multiplying the master equation (\ref{Ex-c}) by $p^n$, and summing over all possible values of $n$ we arrive at a single evolution equation for $G(p,t)$
\begin{equation}
	\label{Ap-c}
	\partial_tG=(p-1)\left\{\left[Bpf(p)-1-\frac{B}{N}\right]G_p-\frac{B}{N}pG_{pp}\right\}.
\end{equation}
Here, we have used the identity $\sum_{n=0}^{\infty}p^n\sum_{k=0}^nD(k)(n-k)\mathbb{P}_{n-k}= \sum_{k=0}^{\infty}D(k)\sum_{n=k}^{\infty}p^n(n-k)\mathbb{P}_{n-k}$ and defined $f(p)=[\sum_{k=0}^{\infty}p^kD(k)-1]/[\langle k\rangle(p-1)]$.

At this point we employ the spectral formalism and expand the solution for $G(p,t)$ in the yet unknown eigenmodes and eigenvalues of the problem. Focusing, however, on times $t\gg t_r$ (where $t_r$ is the relaxation time to the metastable state), and since higher modes only contribute to short-time transients and decay at times on the order of $t_r$, we can write~\cite{spectral_I,spectral_II,spectral_III,MS}
\begin{equation}
	\label{Ap-e}
	G(p,t)\simeq1-\varphi(p)e^{-Et}.
\end{equation}
Here, the stationary solution equals to $1$ corresponding to extinction, $\varphi(p)$ is the lowest excited eigenmode, and $E$ is the lowest excited eigenvalue which equals the inverse of the MTE~\cite{spectral_II,spectral_III,MS}. Note, that this ansatz indicates that $\varphi(0)=1$, to ensure that the probability distribution is normalized at all times.

Plugging Eq.~(\ref{Ap-e}) into (\ref{Ap-c}) we result with a homogenous ordinary differential equation for $\varphi(p)$
\begin{equation}
	\label{Ap-f}
	\frac{B}{N}p(p-1)\varphi''(p) +(p-1)\left[1+\frac{B}{N}-Bpf(p)\right]\varphi'(p) -E\varphi(p) = 0,
\end{equation}
where prime denotes differentiation with respect to $p$. Eq.~(\ref{Ap-f}) has two singular points at $p=0$ and $p=1$ which defines the region of interest. Thus, there are two self-generated boundary conditions $\varphi(1)=0$, and $(1+B/N)\varphi'(0)+E\varphi(0)=0$. Since $E$ turns out to be exponentially small, see below, the latter boundary condition can be approximately written as $\varphi'(0)=0$.

We will solve Eq.~(\ref{Ap-f}) in two distinct regimes. In the bulk, $0\leq p< 1$, where $\varphi(p)$ is almost constant, and in a boundary layer $1-p\ll 1$ where it rapidly decreases to $0$~\cite{spectral_II,spectral_III,MS}.

In the bulk, we look for the solution as $\varphi(p)=1+\delta\varphi(p)$ and denote $u(p)=\varphi'(p)=\delta\varphi'(p)$. Casting the equation into a self-adjoint form and approximating the term $E\varphi(p)\simeq E$~\cite{spectral_II,spectral_III,MS}, we arrive at
\begin{equation}
	\label{Ap-j}
	\left[p \ e^{NS(p)}u(p)\right]'-\frac{N E}{B(p-1)}e^{NS(p)}=0,
\end{equation}
where $S(p)= - \int_1^p[f(p')-1/(Bp')]dp'$. Using the boundary condition $u(0)=0$, the solution of this equation is
\begin{equation}
	\label{Ap-k}
	u^{\mathrm{bulk}}(p)=\frac{N E}{B p}e^{-NS(p)}\int_0^p\frac{e^{NS(p')}}{p'-1}dp',
\end{equation}
which is valid at $0\leq p<1$ as long as $1-p\gg 1/N$.

Next, we find the solution in the boundary layer. Here, as can be checked a-posteriori, since the derivatives of $\varphi(p)$ are large, we can neglect the term $E\varphi$ in Eq.~(\ref{Ap-f}), and arrive at the equation
\begin{equation}
	\label{Ap-k1}
	\frac{B}{N}p(p-1)u'(p) +(p-1)\left[1+\frac{B}{N}-Bpf(p)\right]u(p) = 0.
\end{equation}
What is the boundary condition for $u(p)$ at $p=1$? On the one hand, using Eq.~(\ref{Ap-e}) we have  $\partial_pG(1,t)\simeq-u(1)e^{-Et}$. On the other hand, from Eq.~(\ref{Ap-b}) we have $\partial_pG(1,t)=\bar{n}=n_s e^{-Et}$~\cite{MS}, where $n_s=N(1-1/B)$. Therefore $u(1)=-n_s$, and the solution to Eq.~(\ref{Ap-k1}) reads
\begin{equation}
	\label{Ap-g}
	u^{\mathrm{BL}}(p) = -\frac{n_s}{p}e^{-NS(p)}.
\end{equation}

We now match the bulk and boundary-layer solutions in their joint region of applicability $N^{-1}\ll 1-p\ll 1$. Since $f(p_f)=1/(Bp_f)$, see above, we have $S'(p_f)=0$, and thus, the integrand in Eq.~(\ref{Ap-k}) receives its maximal value at $p_f$. Therefore, when the upper limit of the integral is sufficiently far from $p_f$ we can evaluate the integral in Eq. (\ref{Ap-k}) via the saddle point approximation~\cite{spectral_III}, and arrive at
\begin{equation}
	\label{Ap-m}
	u^{\mathrm{bulk}}(p)\simeq-\frac{N E}{B p(1-p_f)}\sqrt{\frac{2 \pi}{N|S''(p_f)|}}e^{N\left[S(p_f)-S(p)\right]}.
\end{equation}
Here, we have extended the lower and upper limits of the integral to $-\infty$ and $\infty$, respectively, which can be justified a-posteriori using the fact that the width of the Gaussian is much smaller than the distance of $p_f$ to either limits.
Matching this result to the boundary-layer solution [Eq.~(\ref{Ap-g})] we find the MTE $\tau=1/E$ to be
\begin{equation}
	\label{Ap-n}
	\tau=\frac{1}{(B-1)(1-p_f)}\sqrt{\frac{2\pi}{N|S''(p_f)|}}e^{NS(p_f)},
\end{equation}

Let us now recast the MTE (\ref{Ap-n}) in real-space coordinates. The momentum-space coordinates, $q$ and $p$, are related to the real-space coordinates, $Q$ and $P$, via a canonical transformation~\cite{Elgart}
\begin{align}
	\label{Ap-a}
	q & = Qe^{-P}, \nonumber \\
p & =e^P.
\end{align}
Employing this transformation, one can show that $f(p)\to\mathcal{F}(P)$, $p_f\to e^{P_f}$, $S(p_f)\to\mathcal{S}(0)$, and $S''(p_f)\to Q_a'(P_f)e^{-2P_f}$,
where $\mathcal{F}(P)$ is defined below Eq.~(\ref{Ex-f}), $Q_a(P)$ and $\mathcal{S}(0)$ are given by Eqs. (\ref{Ex-g}) and (\ref{Ex-added}), respectively, and  $P_f$ is defined by the relation $Q_a(P_f)=0$. Employing these relations, the MTE given by Eq.~(\ref{Ap-n}) becomes Eq. (\ref{Ex-m}).


\end{document}